# Generalized Charge Energy Rate for Organic Solids and Biomolecular Aggregates Through Drift-Diffusion and Hopping Transport Equations: A Unified Theory

K. Navamani and Swapan K. Pati

*Theoretical Sciences Unit, Jawaharlal Nehru Centre for Advanced Scientific Research*

*Jakkur-560064, Bangalore, India*

**Abstract**

We derive generalized charge energy rate equations for organic solids and biomolecular aggregates, even when these are dynamically disordered. These equations suggest that the transport in such cases rely on both drift and diffusion phenomena. The presence of disorder and field effects makes the equations nonlinear and together with cooperativity, these enhance the charge and energy transport. The generalized drift diffusion expression connects the adiabatic band and nonadiabatic hopping transport mechanisms, well suited for any complex organic semiconductors or assemblies of bio molecular systems. Here we have proposed donor-acceptor (DA) reaction state model, which examines the probability of charge transfer and the rate between two distinct transition state identities. From our analytical equations, we suggest that charge and energy transport property in DA states can be tuned by only a single parameter, i.e., the chemical potential. Importantly, we find the "non-equilibrium assisted drift-diffusion transport" at non-steady state regime in 2D and 3D semiconducting devices. The numerical results clearly support our unified analytical equations, which goes beyond Einstein's diffusion law even in quasi-equilibrium cases.

**Key words**

Charge energy transport, semiconductors, drift-diffusion, disorder, Einstein relation, chemical potential

## I. Introduction

The celebrated Einstein's diffusion-mobility ratio works perfectly for any classical systems. But over the last five decades, there have been many experimental and theoretical results on quantum systems which deviate from the Einstein's classical equation.[1-8] In fact, these deviations are due to the fact that the real materials are not truly classical. In fact, most of these systems are quantum systems, where classical formulations do not work. The quantum systems in general have disorder (strong or weak), defects of various types, electron-electron interactions, electron-phonon interactions etc.[2,3,5-13] These interactions drive the system to reduce symmetries due to energy stability. In such general cases, the classical Einstein equation does not work and importantly it cannot explain the charge transport at low temperature, where most of the phenomena are quantum in nature.[7,8,14,15] The Einstein relation is only valid for non-interacting low density particles systems with equilibrium cases at high temperature.[16] In the context of diode functioning, the Einstein relation very poorly explains the ideality factor for the device performance.[4,12,17] Nowadays, the effect of dynamic disorder is quite important in organic semiconductors, since various inevitable interaction effects, including electronic and nuclear degrees of freedom, controls various parameters like, diffusion, site energy, on-site electronic correlations, local field potential, etc.[9-11,18-24] Commonly, real materials have a number of scattering processes arising from the interactions between charge carriers, with lattice vibrations, with impurities and other carriers etc.. These scattering processes lead the systems to reach nonequilibrium domain, where classical Einstein equation (diffusion-mobility ratio) fail miserably. Recent reports manifest the nonlinearly enhanced electrical transport which further confirms the deviation of Einstein relation, from its classical value of $k_B T/q$.[14,25]

In principle, the charge transport in dynamically disordered systems, like, organic solids and biomolecules is estimated through Master equation method or kinetic Monte-Carlo (MC) simulations.[9,20,23,26-28] The dynamic disorder due to structural kinetics gives rise to on-site potential flux which drifts the carrier motion along the preferred pathway.[20,24,29] In such drift-diffusion cases, the properties can be numerically characterized by the drift disorder time.[21,22,30,31] Here, the drift mobility now takes deviation from the equilibrium mobility (from Einstein relation). Importantly, the dynamical disorder in organic media causes crossover mechanism from hopping to band-like transport.[11,21,22,24,32-35] It is to be noted from various dynamic disorder studies that there is a possibility of band-like transport in organic media due to the dynamic localization,[11,20,21,36] flickering resonance,[33,37] orbital splitting (or degeneracy),[21,38] coherent effect[37,39] and potential induced drift force etc.[7,28,31,32,34,40] In this scenario, both static and dynamic disorder effects need to be included in charge transfer analysis in the multi-site electronic media (multiple local minima of potential energy surfaces) through the key parameters like density flux, diffusion, free energy, polaronic relaxation and disorder-induced potential.[10,11,18,21,22,31,41-43] To analyse the disorder effect on charge transfer kinetics, the entropy is effectively considered here for charge separation and for drift-diffusion studies.[12,41,44,45] It is to be noted that Mendels and Tessler described the charge energy transport during the drift-diffusion process in the degenerate disordered semiconductors.[13] Also, they pointed out that the dependence of charge density on energy transport is an intensive matter in high density devices (degenerate classes of materials), which actually leads to chemical potential dependent drift-diffusion property.

From the discussion above, it is clear that there is not a single complete theory which would explain various diffusion mobility ratio in quantum systems. We attempt to derive a unified theory, taking into account the key issues like, charge-energy transfer rate, charge density, dielectric effect, drift-diffusion and their dependence on disorder for general classes of weak to strong quantum systems, namely, organic molecular solids, semiconducting materials, biomolecular assemblies.

## II. Theoretical Formalism

We have generalized charge-energy rate equations for one dimensional (1D), two dimensional (2D) and three dimensional (3D) organic disordered semiconductors and it can be described as,

$$\left.\frac{dE}{dt}\right|_{1d} = \frac{\pi\hbar e}{2m} n_{1d}\left(\frac{\partial V}{\partial x}\right)_{1d} + \frac{\pi e^2}{24\varepsilon} D_{1d} n_{1d}^3 \tag{1}$$

$$\left.\frac{dE}{dt}\right|_{2d} = \frac{\hbar e(2\pi)^{1/2}}{m} n_{2d}^{1/2}\left(\frac{\partial V}{\partial x}\right)_{2d} + \left(\frac{8}{9\pi}\right)^{1/2} \frac{e^2}{\varepsilon} D_{2d} n_{2d}^{3/2} \tag{2}$$

$$\left.\frac{dE}{dt}\right|_{3d} = \frac{\hbar e}{m}\left(3\pi^2 n_{3d}\right)^{1/3}\left(\frac{\partial V}{\partial x}\right)_{3d} + \frac{e^2}{\varepsilon} D_{3d} n_{3d} \tag{3}$$

where, $\hbar$ is the Planck constant divided by $2\pi$, $e$ is the electronic charge, $m$ is the effective mass of carrier, $n$ is the electron density, $\varepsilon$ is the electric permittivity of the material, $D$ is the diffusion coefficient, and $\partial V/\partial x$ is the gradient of potential along the localized sites. In the above generalized charge-energy rate expressions, the first term on the right side represents the drift type carrier transport and second term represents the diffusive like transport. When there is no field, the charge energy rate expressions purely depend on diffusion property. For homogeneous materials, the energy rate expressions depend on electric field due to negligible effect of density gradient, $(\partial n/\partial x) \to 0$, in all dimensions. Analytically we find that the drift as well as the diffusive transports depend nonlinearly on the disorder, due to the random nature of electronic sites, for any dimensional (1D, 2D and 3D) systems. This randomness of electronic sites takes the system to non-equilibrium situation, which can be controlled by application of electric field, $\vec{E} = \partial V/\partial x$. The correlation between electron density in both equilibrium and nonequilibrium cases due to intrinsic disorder can then be expressed as (for all dimensionality),

$$\left.n_{1d}\right|_{non-equ} = \left.n_{1d}\right|_{equ} \exp\left(\frac{S}{3k_B}\right) \tag{4}$$

$$\left.n_{2d}\right|_{non-equ} = \left.n_{2d}\right|_{equ} \exp\left(\frac{S}{2k_B}\right) \tag{5}$$

$$n_{3d}\big|_{non-equ} = n_{3d}\big|_{equ} \exp\left(\frac{3S}{5k_B}\right) \qquad (6)$$

where, $S$ is the entropy which quantifies the amount of disorder due to randomness and various interactions, and $k_B$ is the Boltzmann constant. The entropy can be estimated by the relation, $S = k_B \ln Z$, where $Z$ accounts for all possible existing electronic states. In principle, the presence of disorder determines the bandwidth and its possible shift influences the electronic transport.[8,44,45] Thus, the degenerate disordered organic semiconductors reflects the large density materials which is responsible for nonlinearly enhanced mobility. Recent numerical simulations confirm the charge transport enhancement in organic semiconductors due to the non-Condon principle, dynamic localization and orbital splitting, etc.[11,20,21,24,30,34] In such degenerate classes of semiconductors, the charge transport deviates from the hopping regime and shows band-like behaviour. This crossover mechanism strongly underlines the nonequilibrium diffusion transport. Generally, the diode functioning can be investigated by the ideality factor $g$ (or enhancement parameter, $g \geq 1$) and with this the modified Einstein relation can be written as,[8,46]

$$\frac{D}{\mu} = \frac{n}{e\frac{\partial n}{\partial \eta}} = g\frac{k_B T}{e} \qquad (7)$$

where, $\eta$ is the chemical potential.

In the present study, we have developed expressions to understand the drift diffusive behaviour in real materials via the enhancement parameter, $g$, which is a function of entropy, temperature, chemical potential and charge density. It has been noted that the parameter $g$ has nonlinear dependence on $n$, chemical potential, $\eta$ and disorder. The main point is that the $g$ value varies linearly with respect to intrinsic disorder for 1D semiconductors, but for other cases (2D and 3D) it has nonlinear dependence on disorder. Interestingly, for 1D semiconductor, the non-exponential behaviour of disorder dependence on hopping diffusion coefficients tends to the linear version of nonanalytic character at low electric fields for all assumed models of disorder, which agree well with the previous theoretical predictions by Nenashev *et al.*[5,6] Now the dimensional effect on *S*, *n*, *T* in diffusion transport can be described as,

$$D_{1d} = g(S=0, T, n, \vec{E}); \quad D_{2,3d} = g(S>0, T, n, \vec{E}); \quad (8)$$

In zero electric field conditions, the diffusion coefficient can be written as,

$$D_{1d}\big|_{non-equ} = D_{1d}\big|_{equ} = f_1(S=0), \quad (9)$$

$$D_{2d}\big|_{non-equ} = D_{2d}\big|_{equ} \exp\left(\frac{S}{4k_B}\right) = f_2(S), \quad (10)$$

$$D_{3d}\big|_{non-equ} = D_{3d}\big|_{equ} \exp\left(\frac{2S}{5k_B}\right) = f_3(S), \quad (11)$$

The above diffusion expressions clearly indicate that there is no intrinsic disorder effect on nonlinear diffusion transport in 1D degenerate systems, which indicates linear and nonanalytic behaviour of *D* at low electric field, $\vec{E}$. It has been noted that the effect of intrinsic disorder on diffusion diminishes while the system size reduces from bulk (3D) to nanoscale (low dimension). Here the proposed equations of entropy dependent diffusion transport for 1D, 2D and 3D systems are in well agreement with an earlier report.[44]

The observed crossover in charge transport equation in various disordered semiconductors asks for the generalization between the localized hopping and the delocalized band transport mechanism.[11,20,22,24,33,34,36] The electronic and nuclear degrees of freedom along with inter-site fluctuations give rise to the polaronic cloud in the materials which effectively determines the electronic transport.[47,48] Notably, dynamically induced dielectric property is commonly observed in most of the disordered (degenerate) semiconductors.[11,47,49] Due to on-site potential flux, there is current density gradient along the hopping sites, and the charge carrier dynamics is polaron dominated.[22,27,43,50] In this work, we generalize the hopping and band transport and write the relaxation time as

$$\tau_{relax} = \left(\frac{\varepsilon m}{ne^2}\right)\frac{\partial P}{\partial t} \qquad (12)$$

where, $\partial P/\partial t$ is the rate of transition probability which is an equivalent of charge transfer rate. Note that, the above generalized hopping-band transport expression purely depends on electric permittivity of the medium ($\varepsilon$), effective mass (m), electron density (n) and transition probability rate. This generalized hopping-band transport equation agrees well with the earlier experimental Hall-effect measurement carried by Podzorov *et al.*[51] and also well-settles with the Trosi's arguments.[11]

In this model, we assume that the electron transfer takes place from donor site to acceptor site (anode to cathode) and hence the chemical potential of donor and acceptor sites are $+\eta_D$ and $-\eta_A$, respectively. Commonly, an electron distribution in the disordered organic semiconductors can be evaluated by the classical Maxwellian form. In such weakly correlated cases, the charge density of donor system can be well-approximated using Boltzmann thermal weightage as,

$$n_D(E_n, \xi_D) = \int_{-\infty}^{\infty} f_{MB}(E_n, \xi_D) g(E) dE \qquad (13)$$

where, $E_n$ is the normalized energy $E_n = E/k_B T$, $\xi_D$ is the normalized chemical potential $\eta_D/k_B T$, $f_{MB}(E_n, \xi_D)$ is the Maxwell-Boltzmann (MB) distribution function, and $g(E)$ is the density of states (DOS) function. In similar way, the charge density of acceptor also can be expressed, $n_A(E_n, \xi_A)$. In principle, the probability of electron transfer from donor to acceptor strongly relies on chemical potentials, $+\eta_D$ and $-\eta_A$. The maximum possibility of electron transfer density ($n_{CT}$) from donor to acceptor states can be calculated directly without any approximation and can be written as,

$$n_{CT} = n_D(E_n, \xi_D) - n_A(E_n, \xi_A) \qquad (14)$$

Here, the probability of charge transfer between donor and acceptor states can be described as,

$$P = 1 - \frac{n_A(E_n, \xi_A)}{n_D(E_n, \xi_D)} \qquad (15)$$

For low density cases, we have developed the probability of charge transfer equation using MB distribution function and it can be expressed as,

$$P = 1 - \exp\left(-\frac{\eta_D + \eta_A}{k_B T}\right) \qquad (16)$$

The above classical description of charge transfer probability is suitable for 1D, 2D and 3D non-interacting particles systems. Here, the chemical potential of acceptor state is equivalent to negative value of the electronegativity, $\eta_A = -\chi_A$. In principle, $\eta_A$ and $\chi_A$ represent donating and accepting tendencies of electrons, which describe the nature of electronic transport property of donor and acceptor states, respectively. The term, $\exp(-(\eta_D - \chi_A)/k_B T)$ represents activation barrier for charge transfer reaction. The calculated CT rate of probability can be written as,

$$\frac{\partial P}{\partial t} = \frac{1}{k_B T}\left(\frac{\partial \eta_D}{\partial t} - \frac{\partial \chi_A}{\partial t}\right)\exp\left(-\frac{\eta_D - \chi_A}{k_B T}\right) \qquad (17)$$

where, $\partial n_D/\partial t$ and $\partial n_A/\partial t$ are the chemical potential flux rate which arise due to dynamical disorder, impurity scattering, applied electric field and lattice interactions, etc. Inserting Equation (17) in generalized hopping-band transport equation (12), results in

$$\tau_{relax} = \left(\frac{\varepsilon m}{n e^2}\right)\frac{1}{k_B T}\left(\frac{\partial \eta_D}{\partial t} - \frac{\partial \chi_A}{\partial t}\right)\exp\left(-\frac{\eta_D - \chi_A}{k_B T}\right) \qquad (18)$$

In crystalline semiconducting, the CT occurs from donor to acceptor and using Fermi-Dirac (FD) distribution function, the maximum probability for CT can be expressed as,

$$P = 1 - \exp\left(-\frac{\eta_D - \chi_A}{k_B T}\right) O(\eta_D, \eta_A, T) \qquad (19)$$

where, $O(\eta_D, \chi_A, T) \approx \left[\dfrac{1 + \exp\left(\dfrac{\eta_D}{k_B T}\right)}{1 + \exp\left(\dfrac{\chi_A}{k_B T}\right)}\right].$

The function $O$ actually accounts for quantum corrections for degenerate semiconductors. For high density limit, the rate of probability of CT can be expressed as,

$$\frac{\partial P}{\partial t} = \exp\left(-\frac{\eta_D - \chi_A}{k_B T}\right)\left[O(\eta_D, \chi_A, T)\frac{1}{k_B T}\left(\frac{\partial \eta_D}{\partial t} - \frac{\partial \chi_A}{\partial t}\right) - \frac{\partial O(\eta_D, \chi_A, T)}{\partial t}\right] \qquad (20)$$

By using Equation (20), the generalized hopping-band transport equation (12) for strongly interacting semiconducting materials can be modified as,

$$\tau_{relax} = \left(\frac{\varepsilon m}{n e^2}\right)\exp\left(-\frac{\eta_D - \chi_A}{k_B T}\right)\left[O(\eta_D, \chi_A, T)\frac{1}{k_B T}\left(\frac{\partial \eta_D}{\partial t} - \frac{\partial \chi_A}{\partial t}\right) - \frac{\partial O(\eta_D, \chi_A, T)}{\partial t}\right] \qquad (21)$$

From the above equations, it is clear that the CT rate for both low and high charge density systems is directly related to the energy transfer rate, which again is related to the effective electro-chemical potential energy of donor and acceptor states. Using this model, we are able to describe the charge and energy transport in various distinct transition states in different domains like, adiabatic and non-adiabatic. Here derived equations can also be used by experimentalists to design their experiments to improve the device performance as well as stability with the aid of chemical and electrochemical doping (including substituents), applying gate voltage, making supramolecular order, processing, temperature, etc.[52]

To obtain further insight, we consider the effect of disorder, which is present in almost all real materials. According to disorder dependent energy dispersion relation which is proposed in our earlier study (see Eqn. 15 in Ref. 21), one is able to develop the charge transport in degenerate disordered materials, which can be expressed as,[21]

$$\Delta E_S \approx \Delta E_{S=0} \exp\left(-\frac{S}{2k_B}\right) \qquad (22)$$

where, $\Delta E_S$ is the energy gap with disorder. The $S$ is entropy which is the measure of disorder, $S = k_B \ln Z$, with $Z$ as the total number of degenerate states. We can modify it to obtain energies corresponding to bottom of the conduction band $(E_C)$ and top of the valence band $(E_V)$ with respect to chemical potential $(\eta)$ for hole and electron transport,

$$\left. \begin{array}{l} (\eta - E)_S \approx (\eta - E)_{S=0} \exp\left(-\dfrac{S}{2k_B}\right) \\[2mm] (E - \eta)_S \approx (E - \eta)_{S=0} \exp\left(-\dfrac{S}{2k_B}\right) \end{array} \right\| \qquad (23)$$

The contribution of disorder in terms of its weight should be included in the MB and FD distribution functions to estimate the charge transport in the disordered semiconductors (or degenerate materials), via Equations (14) to (21) for weakly and strongly correlated semiconducting materials.

In non-equilibrium cases, the potential drift can be random over sites in degenerate organic systems, which gives drift like transport along the hopping sites,

$$V_d(t) = g(S, \vec{E}, T) \frac{k_B T}{e} \left(1 - (\exp(1 - P(t)))^{2/5}\right) \qquad (24)$$

where, $V_d(t)$ is the potential drift, $g$ is a function of disorder, electric field and temperature, and $P(t)$ represents the survival probability of charge in an initial electronic site. The basic point is that $g$ is always greater than or equal to 1 and it increases drastically with electric field.

We find that the drift potential expression actually dictates the $D/\mu$ values in high electric field (when it becomes nonlinear), where the $g$ parameter is quite large. At weakly disordered (or quasi-equilibrium) and very low field or zero field conditions, the parameter $g \to 1$ in which the drift like transport Equation (24) can be reduced to (see Eqn. 15 in Ref. 31),

$$V_d(t) = \frac{k_B T}{e}\left(1 - \left(\exp(1 - P(t))\right)^{2/5}\right) \qquad (25)$$

Thus, the field enhanced diffusion-mobility equation can be written as,

$$\left(\frac{D}{\mu}\right)_{quasi-equ} = \left(\frac{D}{\mu}\right)_{equ} + V_d(t) \qquad (26)$$

Here, the enhancement parameter ($g$) strongly depends on drift potential and it can be tuned by electric field, disorder and chemical potential. In quasi-nondegenerate materials, the effective diffusion along the localized sites (the shallow potentials) is approximated as, $(D/\mu)_{eff} = (D/\mu)_{equ}\left(1 - \left(\exp(1 - P(t))\right)^{2/5}\right)$. It is clear from this equation that even in quasi-equilibrium cases, the diffusion-mobility ratio deviates from Einstein equation.

## III. Analysis and Applications

Using our formalism, one can clearly predict the drift-diffusion behaviour for any dimensional devices with a variety of intrinsic and external factors, like, weak to strong disorder, low, intermediate to high field effect and linear to nonlinear regime transport behaviours.[43,44,47,48,53] Interestingly, various earlier reports suggested the entropy effect on charge separation in organic photovoltaic cells in different dimensional systems (1D, 2D and 3D).[44,45] Here the Coulomb barrier is a key factor for electronic transport via charge separation, and is strongly influenced by entropy (or degeneracy) for 2D and 3D materials.[44] But in the case 1D, there is no entropy role for the electronic transport. These observations are clearly shown in our entropy dependent diffusion formalism for different dimensional cases (see Eqns. 9, 10 and 11). Recent experimental investigations explicitly provide the hole transport in three different organic thin film devices, **P1**, **P2** and **P3** at different bias voltages.[50] These devices originally fabricated by using alkyl-substituted Triphenylamine (TPA) based molecules such as, 2-(4-(5-(4-(diphenylamino)phenyl)-1,3,4-oxadiazol-2-yl)benzylidene)malononitrile **($X_1$)**, 2-(4-(5-(4-(di-p-tolylamino)phenyl)-1,3,4-oxadiazol-2-yl)benzylidene)malononitrile **($X_2$)** and 2-(4-(5-(4-(bis(4-(tert-butyl)phenyl)amino)phenyl)-1,3,4-oxadiazol-2-yl)benzylidene) malononitrile **($X_3$),** respectively. It has been reported that the electronic device, **P3,** has large hole transporting ability in comparison to **P2** and **P1** devices**.**

To estimate the non-equilibrium effect on charge transport in **P3** device, we have numerically performed survival probability of the charge carrier at initial localized site, and diffusion behaviour with respect to the time, with the aid of Monte Carlo (MC) simulation. The survival probability plot (see Fig. 1) provides the charge decay nature of P3 device, which can be controlled by bias voltage, doping, disorder and other dynamical effects. Notably, the cooperativity behaviour of non-equilibrium and equilibrium charge transport are found in case of P3 device, which leads mainly to the drift-diffusion mobility (see Fig. 2) which explain the higher carrier transporting ability of P3 device. However, in the case of P1 and P2 devices, the cooperativity behaviour is minimal.

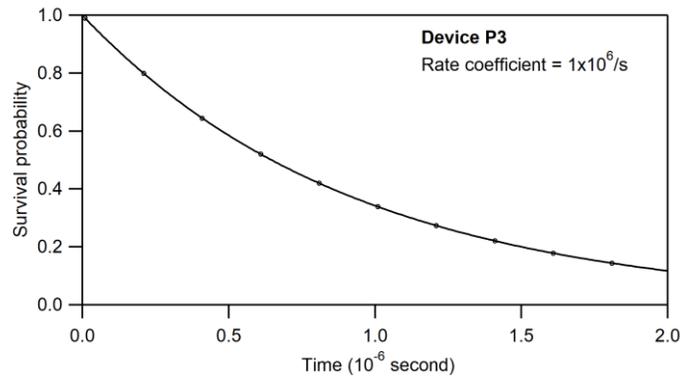

FIG. 1. Survival probability of hole carrier in P3 device

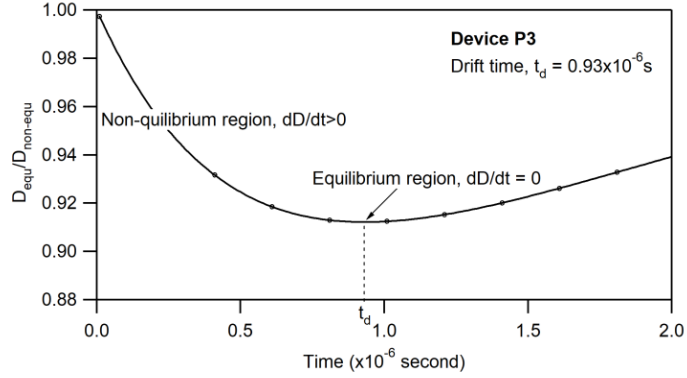

FIG. 2. Cooperative behaviour of both equilibrium and non-equilibrium diffusive transport at steady and non-steady state regimes in P3 device

There is a relationship between drift-diffusion and non-steady state, $\frac{dD}{dt} \neq 0$, as well as steady state, $\frac{dD}{dt} = 0$, which is related with the density flux behaviour. We find that the density flux due to dynamical disorder facilitates the non-equilibrium diffusive transport, due to which, it deviates from Einstein's classical original value of $\frac{k_B T}{e}$. Interestingly, the disorder drift time acts as the crossover point between the non-equilibrium and equilibrium transport, which was also reported earlier.[22,31,35] From Fig. 2, we can relate the equilibrium and non-equilibrium diffusive transport and it can be expressed as,

$$\frac{dD}{dt} \equiv \frac{D_{non-equ} - D_{equ}}{t_d}, \qquad (27)$$

where $t_d$ is the disorder drift time. We can rewrite the above Equation as

$$D_{non-equ} = \frac{dD}{dt} t_d + D_{equ} \qquad (28)$$

Inserting Equation (28) in Einstein's diffusion-mobility relation gives us

$$\left(\frac{D}{\mu}\right)_{non-equ} = \frac{1}{\mu}\left(\frac{dD}{dt}t_d\right) + \left(\frac{D}{\mu}\right)_{equ} \quad (29)$$

The first term of right side in the above Equation (29) is to be termed as drift kind diffusive mobility, which is equivalent to the drift potential, $V_{drift}$. The second term is related to the classical Einstein's equation for pure diffusive transport (zero drift) and is normally observed at equilibrium domain. The above equation (Eq. 29) is to be expected for the "non-equilibrium assisted drift-diffusion transport" at non-steady state regime in 2D and 3D organic semiconducting devices. In steady state limit, this equation is reduced to the original Einstein relation, $\frac{D}{\mu} = \frac{k_B T}{e}$.

To get further insight on drift-diffusion cooperative behaviour, through our generalized equations, we hereby also extend previous theoretical charge transfer kinetic investigation of octupolar derivatives.[22] In our numerical study, octupolar derivatives considered for charge carrier dynamical analysis are: 2,4,6-tris[5-(3,4,6-trioctyloxyphenyl)thiophene-2-yl]-1,3,5-triazene (octupolar 1b), 2,4,6-tris[5-(3,4,6-trimethoxyphenyl)thiophene-2-yl]-1,3,5-triazene (octupolar 1c) and 2,4,6-tris[5′-(3,4,6-tridodecyloxyphenyl)-2,2′-bithiophene-5-yl]-1,3,5-triazene (octupolar 2), which were synthesized by Yasuda *et al*.[56] Here, we have included the structural dynamic effect on our generalized charge transport model through MC simulation in three different systems, octupolar 1b, octupolar 1c and octupolar 2.

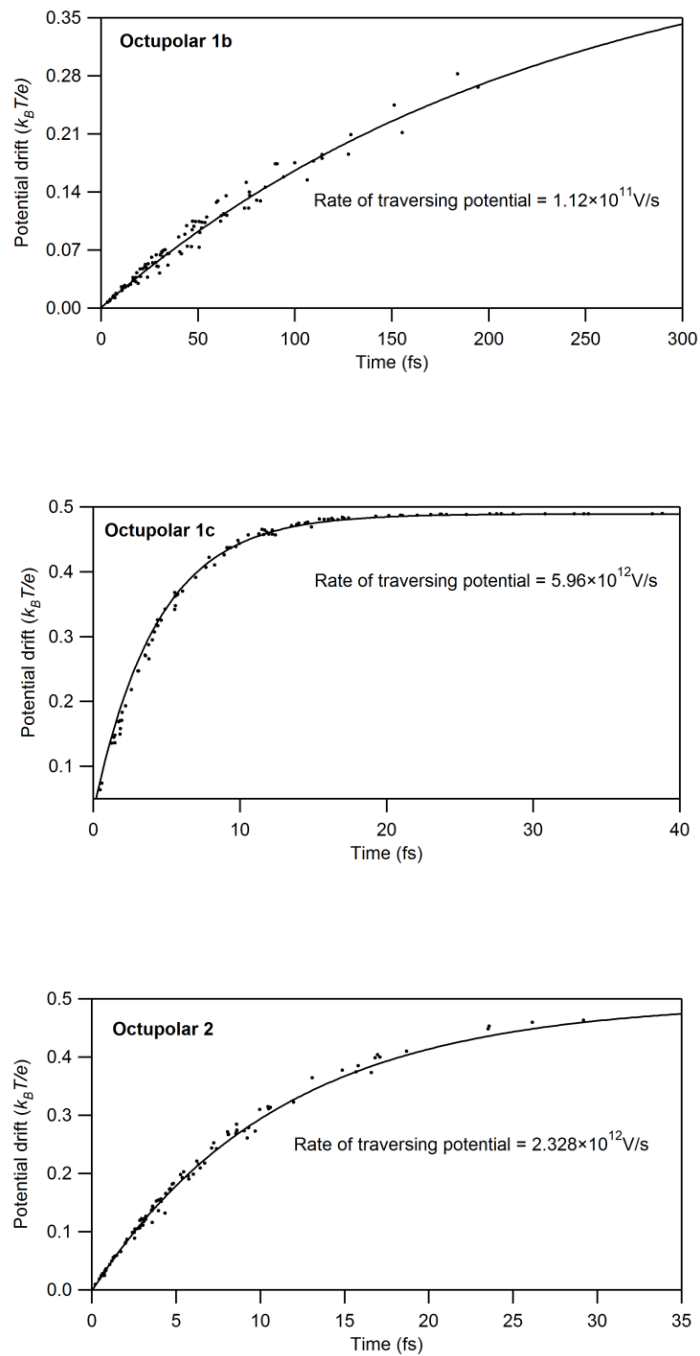

**FIG. 3.** The traversing potential along the electronic sites during the simulation for charge transport in octupolar derivatives. The potential flux takes over from the linear transport, facilitates nonlinear behaviour at any degenerate cases even in quasi-equilibrium situation.

The performed MC simulation yields time gap between nonequilibrium and equilibrium charge transfer process by disorder, even at zero field, also we monitored the crossover charge transfer scenario (due to random on-site potential fluctuation). In such disordered cases, required field induced potential drift for charge carrier motion along the preferred charge transfer path (along the localized sites/conjugated length) is numerically calculated and is shown in Fig. 3. It is to be noted that there is an increment on deviation from equilibrium transport (or classical Einstein's diffusion law), which strongly depends on torsional disorder between adjacent units, on conjugation length and on the structural dynamics effect. Numerically obtained parameter 'rate of traversing potential' (see Fig.3) determines the time gap between nonequilibrium and equilibrium transport, essentially depends on onsite-potential flux by dynamic disorder or field. This observation assures the nonlinear behaviour in drift-diffusion process and is well settled with our generalized band-hopping analytic models. Here, the traversing potential rate associated with non- steady state electronic transition and is evaluated by chemical potential flux rate due to interaction between electronic and nuclear degrees of freedom. In this way, local and nonlocal interactions define an intrinsic charge localization strength which quantifies the value of charge transfer density, well suited with an earlier study by Wang and Beljonne.[35] It has been noted that the derivative octupolar 1c has high rate of traversing potential of around $6 \times 10^{12} V/s$ for electron transport which facilitates coherent band like transport and is in agreement with an earlier observation.[22] But in the case of electron transport in octupolar 1b has low traversing potential rate $(1.1 \times 10^{11} V/s)$ which stipulates less density flux and is responsible for incoherent hopping transport. This is well suited with a diffusion limited by disorder.[11,18,20,36] The simulation results confirm that the accuracy of our analytic expressions which basically correlates both the linear and the nonlinear transport phenomena, depends on field, disorder and chemical potential. In this study, we also have confirmed earlier descriptions in appropriate limits, such as, Troisi's localization transport limitation by dynamic disorder,[11,20,36] Tessler's charge energy transport in degenerate organic semiconductors,[12,13] Gregg's entropy dependent charge separation process[44] and Beljonne's crossover transport mechanism.[35]

## IV. Conclusions

The proposed charge energy transfers rate expressions for 1D, 2D and 3D semiconducting devices depend on both drift and diffusion transport, which are well-parameterized by the effective mass, charge density, net electric field, dielectric constant and diffusion coefficient. The derived general drift-diffusion equations show the crossover between hopping and band transport characteristics. Interestingly, it may show either linear or nonlinear behaviour, depending on the extent of dynamic disorder and potential drift. The developed charge transfer equation on the basis of donor-acceptor state model quantifies the maximum probability of charge transfer and the corresponding rate describes two distinct electronic sites, in both weakly and strongly interacting materials. The nonlinearity of drift-diffusive electronic transport reveals the limitations of Einstein equation even at quasi-equilibrium. Our unified theory mainly suggests the "non-equilibrium assisted drift-diffusion transport" at non-steady state situations for 2D and 3D semiconducting devices, and in some appropriate limits, validates a few earlier reports. Using potential drift equation, the average dropping potential by disorder can also be numerically calculated for any organic semiconducting devices or biomolecular assemblies.